\documentclass{epl}

\usepackage{amssymb}
\usepackage{amsfonts}
\title{Heat fluctuations for harmonic oscillators}
\author{S. Joubaud, N. B. Garnier, S. Ciliberto}
\institute{Laboratoire de Physique de l'ENS Lyon, CNRS UMR 5672,
        46, All\'ee d'Italie, 69364 Lyon CEDEX 07, France}
\date{\today}

\begin{document}

\maketitle

\begin{abstract}
Heat fluctuations of a harmonic oscillator in contact with a thermostat
and driven out of equilibrium by an external deterministic force are studied experimentally
and theoretically within the context of Fluctuation Theorems.
We consider the case of a periodic forcing of the oscillator, and we calculate
the analytic probability density function of heat fluctuations.
The limit of large time is discussed and we show that heat fluctuations
satisfy the conventional fluctuation theorem, even if a different fluctuation
relation exists for this quantity. Experimental results are also given
for a transient state.

\end{abstract}

Out-of-equilibrium fluctuations have recently received a lot of attention,
especially in the context of nanosystems and biomolecules where fluctuations
are large. In the last decade, Fluctuation Theorems (FTs) appeared in
nonequilibrium physics. They quantify the asymmetry of fluctuations of
entropy production for a large class of systems, possibly far from equilibrium.
These theorems were first demonstrated
in deterministic dynamics of many degrees of
freedom~\cite{Evansetal93,GallavottiCohen95a} and later extended to stochastic
systems~\cite{Kurchan98,Farago,Cohen,Cohen1}. The FT states that the probability $P(\sigma = a)$
of observing an entropy production rate $\sigma$, measured over a time
$\tau$, with a value $a$, satisfies
\begin{equation}
\frac{P(\sigma=a)}{P(\sigma=-a)}  \rightarrow  \exp (a \tau) \quad {\rm for \,\, large} \,\, \tau \,\, {\rm and \,\, any \,\, value} \,\, a
\label{FT}
\end{equation}
There are two classes of FTs. The {\em Stationary State Fluctuation Theorem}
(SSFT) considers a nonequilibrium stationary state. SSFT holds in the limit
of large time $\tau$. The  {\em Transient Fluctuation Theorem} (TFT) describes transient nonequilibrium states
where $\tau$ measures the time interval since the system left its equilibrium state.
Contrary to SSFT, the TFT holds for all times, {\em i.e.}, equation~(\ref{FT}) is an equality even for
arbitrarily small values of $\tau$.

In this letter, we consider the heat $Q_\tau$ dissipated on a time interval $\tau$
in a thermostated system at temperature $T$ rather than the entropy production.
We study the fluctuations of the heat dissipated by a harmonic oscillator in
contact with the thermostat and driven out of equilibrium by an external force.
Experimental tests of FTs are rare. Some of them are performed in
dynamical systems~\cite{otherexperiment}, in which the interpretation of the
results is very difficult. In stochastic systems, a laboratory experiment was
carried out by Wang {\it et al} using a Brownian particle in a moving optical
trap~\cite{Wangetal:02:05}. Work fluctuations were shown to obey the predictions of ref.~\cite{Cohen1}.
Work and heat fluctuations were also studied in an electrical circuit by Garnier and Ciliberto~\cite{Garnier} ;
the theoretical predictions for both heat and work fluctuations were satisfied~\cite{Cohen1, Cohen}.
These two systems are described by a first order Langevin equation. Systems described by a second order Langevin equation have been studied~\cite{Kurchan98}, and tested experimentally for work fluctuations $W_\tau$~\cite{Douarche2006}; as far as we know,
no analytical results for the probability density functions (PDFs) of heat
fluctuations $Q_\tau$ are available in this case. In ref.~\cite{Cohen1},
van Zon and Cohen have calculated the Fourier transform of the PDF of
$Q_\tau$ for a first order Langevin dynamics but no exact expression of the
PDF itself is known.

In the following, we consider first a stationary state obtained by driving
the system periodically in time. We calculate exactly the probability density
function of $Q_\tau$. We then compare our finding with new experimental results,
and show that SSFT holds, and we discuss the large time limit.
Finally experimental results of a TFT for the heat are reported.

To study FT, we measure the out-of-equilibrium fluctuations of a
harmonic oscillator whose damping is due to the viscosity of a
surrounding fluid, acting as a thermal bath. The oscillator is a
torsion pendulum composed of a brass wire and a glass mirror glued
in the middle of this wire. The elastic torsional stiffness of the
wire is $C = 4.7 \times 10^{-4}$ $\textrm{N\,m\,rad}^{-1}$. It is
enclosed in a cell filled by a water-glycerol solution at $60 \%$
concentration. The system is a harmonic oscillator with resonant
frequency $f_o=\sqrt{C/I_{\rm eff}}/(2\pi)=217$ Hz and a
relaxation time $\tau_\alpha=2I_{\rm eff}/\nu= 9.5$ ms.
$I_{\mathrm{eff}}$ is the total moment of inertia of the displaced
masses and $\nu$ is the oscillator viscous damping. The angular
displacement of the pendulum $\theta$ is measured by a
differential interferometer~\cite{rsi, DouarcheJSM}. The
measurement noise is two orders of magnitude smaller than thermal
fluctuations of the pendulum. $\theta(t)$ is acquired with a
resolution of 24 bits at a sampling rate of $8192$ Hz, which is
about 40 times $f_o$. The calibration accuracy of the apparatus,
tested using the Fluctuation Dissipation Theorem, is
better than $3\%$ (see~\cite{DouarcheJSM}). We drive the system
out-of-equilibrium by forcing the system with an external torque
$M$ by means of a small electric current $J$ flowing in a coil
glued behind the mirror. The coil is inside a static magnetic
field, hence $M\propto J$. More details on the set-up can be found
in ref.~\cite{rsi, DouarcheJSM}. The system is very well described
by a second order Langevin equation:
\begin{equation}
I_{\mathrm{eff}}\,\frac{{\rm d}^2{\theta}}{{\rm d}t^2}
+ \nu \,\frac{{\rm d}{\theta}}{{\rm d}t}  + C\,\theta =
M + \sqrt{2k_B T\nu} \, \eta, \label{eqoscillator}
\end{equation}
with $\eta$ the noise, delta-correlated in time,
$\beta = (k_B T)^{-1}$ with $k_B$ Boltzmann's constant and
$T$ the temperature of the system. At equilibrium ($M=0$ pN.m),
the PDF of the thermal fluctuations $\delta \theta$ is Gaussian
with variance $k_B T/C$. We apply two kinds of forcing.
First, we study a periodic forcing : $M(t) = M_o \sin(\omega_d t)$
with $M_o= 0.78$ pN.m and $\omega_d/2\pi=64$ Hz.
We then analyze a linear forcing $M(t) = M_o t/\tau_r$
with $M_o=6.2$ pN.m and $\tau_r$ = 0.01 s = 1.07 $\tau_\alpha$.

The change in internal energy $\Delta U_\tau$ of the oscillator
over a time $\tau$, starting at a time $t_i$, is written as:
\begin{equation}
\Delta U_\tau = U(t_i+\tau) - U(t_i) = Q_\tau + W_\tau
\label{Energyconservation}
\end{equation}
which is the first law of thermodynamics. $W_\tau$ is the work done on the system over a time $\tau$:
\begin{equation}
W_\tau = {1 \over k_B \ T} \int_{t_i}^{t_i+\tau} M(t') \frac{\rm d \theta}{\rm dt}(t') dt'
\label{Wdef}
\end{equation}
and $Q_\tau$ is the heat given to the system.
Equivalently, $(-Q_\tau)$ is the heat dissipated by the system.

For a harmonic oscillator described by a second order Langevin equation,
the internal energy has two contributions : the kinetic and potential energies:
\begin{equation}
U(t)={1 \over k_B \ T} \left[{1 \over 2} I_{\mathrm{eff}}\left[
\frac{\rm d \theta (t)}{\rm dt} \right]^{2} +{1 \over 2} C
\theta(t)^2 \right] \label{Udef}
\end{equation}
Multiplying eq.~(\ref{eqoscillator}) by $\frac{\rm d \theta}{\rm dt}$
and integrating between $t_i$ and $t_i+\tau$, we obtain exactly
the first law of thermodynamics eq.~(\ref{Energyconservation})
and have the following expression for the heat:
\begin{equation}
Q_\tau = \Delta U_\tau - W_\tau = -{1 \over k_B \ T }
\int_{t_i}^{t_i+\tau} \nu \left[ \frac{\rm d \theta}{\rm dt}
(t')\right]^{2}dt' + {1 \over k_B \ T } \int_{t_i}^{t_i+\tau}
\eta(t') \frac{\rm d \theta}{\rm dt}(t') dt' \,. \label{Qdef}
\end{equation}
The first term corresponds to the viscous dissipation and is always positive,
whereas the second term can be interpreted as the work of the thermal noise which have a fluctuating sign.

We rescale the work $W_\tau$ (the heat $Q_\tau$) by the average
work $\langle W_\tau \rangle$ (the average heat $\langle Q_\tau
\rangle$) and define: $w_\tau = \frac{W_\tau}{\langle W_\tau
\rangle}$ ($q_\tau = \frac{Q_\tau}{\langle Q_\tau \rangle}$).
Averages are obtained experimentally as time-averages, and they are proportional to $\tau$ on
the stationary state under consideration.
\begin{figure}
\centerline{\includegraphics[width=0.7\linewidth]{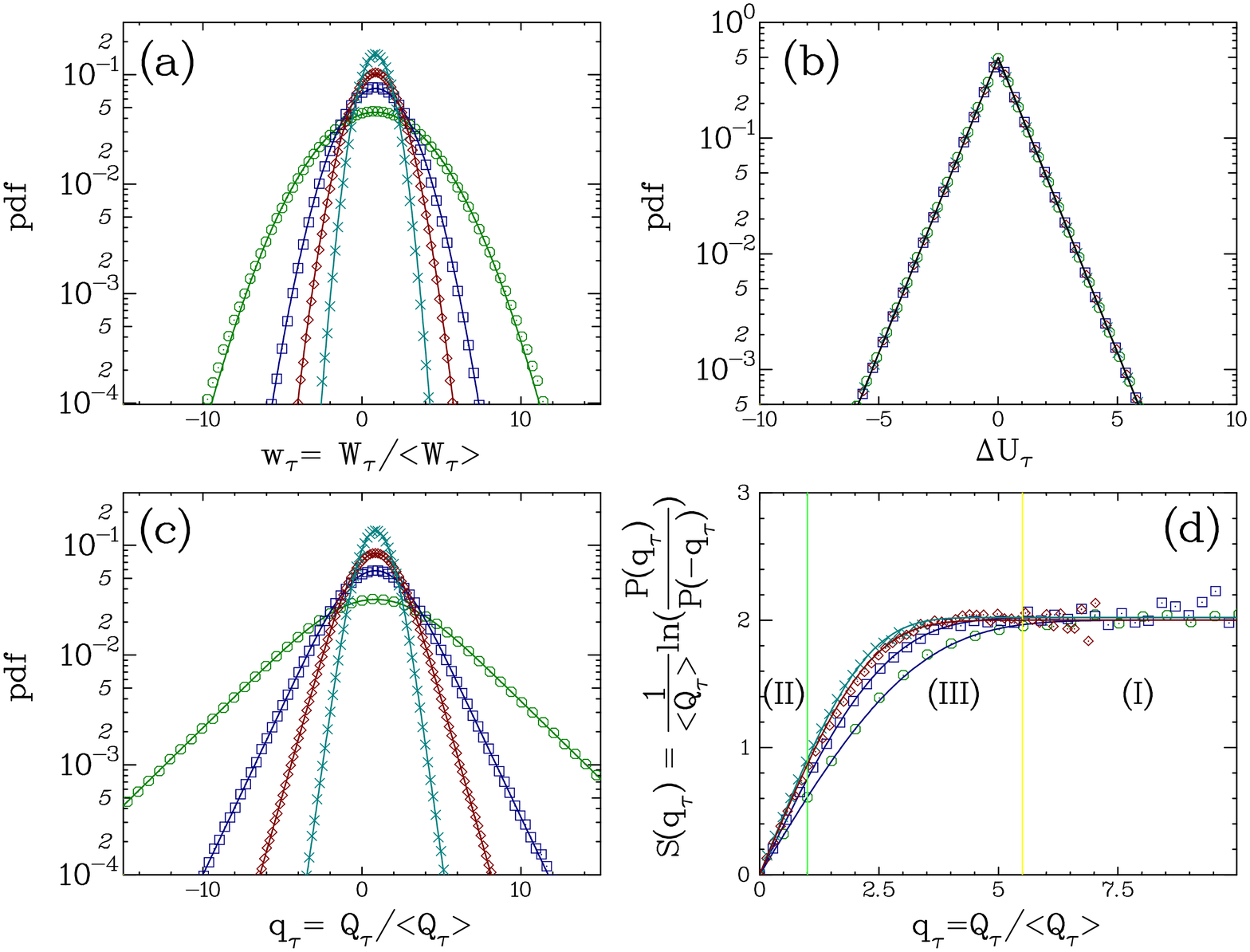}}
\centerline{\includegraphics[width=0.7\linewidth]{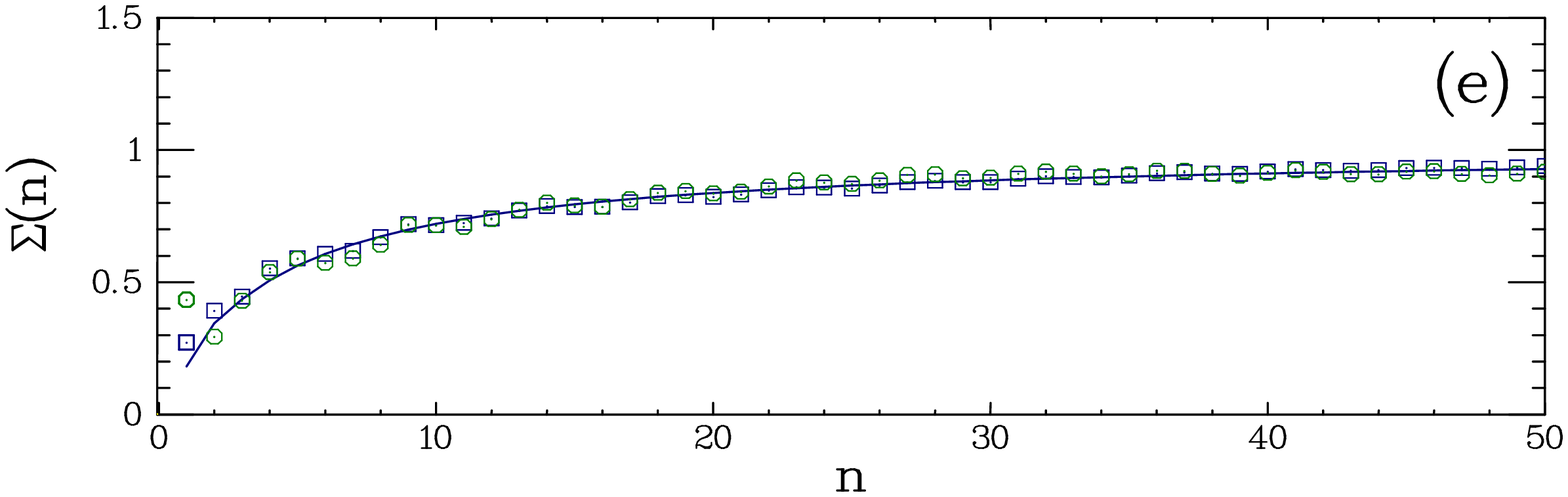}}
\caption{Sinusoidal forcing. a) PDFs of $w_\tau$. b) PDFs of $\Delta U_\tau$. c) PDFs of $q_\tau$.
d) Functions $S(q_\tau)$. In all plots, the integration time $\tau$ is a multiple of the period of forcing, $\tau = 2 n \pi/\omega_d$, with $n=7$ ($\circ$), $n=15$ ($\Box$), $n=25$ ($\diamond$) and $n=50$ ($\times$). Continuous lines are theoretical predictions with no adjustable parameter.  e) The slope $\Sigma_q(n)$ of $S(q_\tau)$ for $q_\tau <1$, plotted as a function of $n$ ($\circ$). The slope $\Sigma_w(n)$ of $S(w_\tau)$ plotted as a function of $n$($\Box$). Continuous line is theoretical prediction.}
\label{fig:sinusFT}
\end{figure}

We consider first the periodic forcing. In this case, we choose
$\tau$ as a multiple of the period of the driving ($\tau = 2 n
\pi/\omega_d$ with $n$ integer). The starting phase $t_i \omega_d$
is averaged over all possible $t_i$ to increase statistics. The
PDFs of $w_\tau$, $\Delta U_\tau$ and $q_\tau$ are plotted in
Fig.~\ref{fig:sinusFT} for different values of $n$. The average of
$\Delta U_\tau$ is clearly zero because the time $\tau$ is a
multiple of the period of the forcing. The PDFs of the work
(fig.~~\ref{fig:sinusFT}a) are Gaussian for any $n$ whereas the
PDFs of heat fluctuations $q_\tau$ have exponential
tails (fig.~~\ref{fig:sinusFT}c). These exponential PDFs can be
understood noticing that, from eq.~(\ref{Qdef}),
$-Q_\tau=W_\tau-\Delta U_\tau$ and that $\Delta U_\tau$ has an
exponential PDF independent of n (fig.~\ref{fig:sinusFT}b).
Therefore, on a first approximation, the PDF of $q_\tau$ is the
convolution between an exponential and a Gaussian.

To quantify the symmetry of the PDF around the origin, we define the function $S$ as:
\begin{equation}
S(e_\tau) \equiv  {1 \over \langle E_\tau \rangle}\ln\left[{P(e_\tau)\over P(-e_\tau)}\right] \label{eq_FT}
\end{equation}
where $e_\tau$ stands for either $w_\tau$ or $q_\tau$ and $E_\tau$ stands for either $W_\tau$ or $Q_\tau$. The question we ask is whether:
\begin{equation}
\lim_{\tau \rightarrow \infty} S(e_\tau) = e_\tau
\label{conventionalFT}
\end{equation}
as required by SSFT. $S(q_\tau)$ is plotted in Fig.~\ref{fig:sinusFT}d for different values of $n$ ; three regions appear:

(I) For large fluctuations $q_\tau$, $S(q_\tau)$ equals $2$.
When $\tau$ tends to infinity, this region spans from $q_\tau = 3$ to infinity.

(II) For small fluctuations $q_\tau$, $S(q_\tau)$ is a linear function of $q_\tau$.
We then define $\Sigma_q(n)$ as the slope of the function $S(q_\tau)$, {\it i.e.} $S(q_\tau) = \Sigma_q(n) \, q_\tau$. This slope is plotted in Fig.~\ref{fig:sinusFT}e where we see that it tends to $1$ when $\tau$ is increased.
So, SSFT holds in this region II which spans from $q_\tau=0$ up to $q_\tau = 1$ for large $\tau$.

(III) A smooth connection between the two behaviors.

The PDF of the work being Gaussian, the functions $S(w_\tau)$ are proportional
to $w_\tau$ for any $\tau$, {\it i.e.} $S(w_\tau)=\Sigma_w(n) \, w_\tau$ (ref.~\cite{Douarche2006}). $\Sigma_w(n)$ is plotted in Fig.~\ref{fig:sinusFT}e and we observe that it matches experimentally $\Sigma_q(n)$, for all values of $n$. So the finite time corrections to the FT for the heat are the same than the ones of FT for work~\cite{Douarche2006} : $\Sigma_w(n)=\Sigma_q(n)=1+K/n+1/n{\cal O} \left({\rm e}^{-\tau_n/\tau_\alpha}\right)$, where $K$ is a constant.

We now give an analytical expression of the PDF of heat $Q_\tau$. In order to do this, we write $\theta = \bar{\theta}+ \delta \theta$ where $\bar{\theta}$ is the mean response of the system to the external work and $\delta \theta$ the thermal fluctuations. We suppose that fluctuations at $M(t) \neq 0$ are those of equilibrium, {\it i.e.} that the external driving does not perturb the equilibrium PDF. This hypothesis is supported experimentally as shown in ref.~\cite{DouarcheJSM} and detailed in~\cite{Douarche2006}. Times $\tau$ under consideration are taken larger than $\tau_\alpha = 9,5 $ ms, so that exponential corrections to the autocorrelation functions, which are scaling like $\exp(-\tau/\tau_\alpha)$, can be neglected. Experimentally, $\tau/\tau_\alpha=1.64\,n$, so this is correct as soon as $n$ is larger than 3 or 4. Within this assumption, $\theta(t_i+\tau)$ and $\theta(t_i)$ are independent,
and so are $\frac{{\rm d}{\theta}}{{\rm d}t}(t_i+\tau)$ and $\frac{{\rm d}{\theta}}{{\rm d}t}(t_i)$.
Eq.~(\ref{eqoscillator}) is second order in time, so $\theta(t)$ and $\frac{{\rm d}{\theta}}{{\rm d}t}(t)$
are independent at any given time $t$. Just like in our experiment, we choose the integration time $\tau$
to be a multiple of the period of the forcing, so
$\langle \Delta U_\tau \rangle = 0$ and therefore
$\langle W_\tau \rangle = - \langle Q_\tau \rangle$.
Within this framework, we find that the PDF of $\Delta U_\tau$ is exponential:
\begin{equation}
P(\Delta U_\tau) = {1 \over 2} \exp (-|\Delta U_\tau|) \,.
\label{PDFdE}
\end{equation}
It is independent of $\tau$ because $\Delta U_\tau$ depends only on $\theta$ and $\frac{{\rm d}{\theta}}{{\rm d}t}$ at times $t_i$ and $t_i+\tau$ which are uncorrelated.
This expression is in perfect agreement with the experimental PDFs for all times (see Fig.~\ref{fig:sinusFT}b). Experimentally, work fluctuations have a Gaussian distribution so it is fully characterized by its mean $\langle W_\tau \rangle$ and its variance $\sigma_W^2$.  In a Gaussian case, the FTs take a simple form $S(w_\tau)= \frac{2 \langle W_\tau \rangle}{\sigma_W^2} w_\tau = \Sigma_w(n) w_\tau$. We have already computed the analytic expression of $\Sigma_w(n)$~\cite{Douarche2006} and this gives a relation between the variance and the mean value of $W_\tau$: $\sigma_W^2 = 2 \langle W_\tau \rangle + {\cal O}({1 \over \tau})$ and so $\sigma_W^2 = 2|\langle Q_\tau \rangle| + {\cal O}({1 \over \tau})$.

To obtain the PDF $P(Q_\tau)$ of the heat, we define its Fourier transform (characteristic function) as
\begin{equation}
\hat{P} _{\tau} (s) \equiv \int_{-\infty}^{\infty}  {\rm d} q_\tau e^{isq_\tau} P(q_\tau)
\label{DefFourier}
\end{equation}
which can be computed exactly~\cite{Cohen1}. We then write $P(q_\tau)$ using eq.~(\ref{Qdef}) as:
\begin{equation}
P(q_\tau) = \int \int {\rm d}\theta {\rm d}\dot{\theta} \tilde{P}\left(\Delta U_\tau-Q_\tau, \theta(t_i+\tau) , \theta(t_i),\dot{\theta}(t_i+\tau), \dot{\theta}(t_i)\right)
\end{equation}
where $\tilde{P}$ is the joint distribution of the work $W_\tau$,
$\theta$ and $\frac{{\rm d}{\theta}}{{\rm d}t}$ at the beginning
and at the end of the time interval $\tau$. This distribution is
expected to be Gaussian because $W_\tau$ is linear in
$\dot{\theta}$ and additionally $\theta$, $\dot{\theta}$ and
$W_\tau$ are  Gaussian. Some algebra then yields:
\begin{equation}
\hat{P} _{\tau} (s)= \frac{1}{1+s^2}\exp \left( i\langle Q_\tau \rangle s-\frac{\sigma_W^2}{2} s^2 \right)
\label{TFPDFQ}
\end{equation}
The characteristic function of heat fluctuations
is therefore the product of the characteristic function of an exponential distribution ($\frac{1}{1+s^2}$)
with the one of a Gaussian distribution ($\exp \left( i\langle Q_\tau \rangle s-\frac{\sigma_W^2}{2} s^2 \right)$).
Thus the PDF of heat fluctuations is nothing but the convolution of a Gaussian and an exponential PDF,
just as if $W_\tau$ and $\Delta U_\tau$ were independent. The inverse Fourier transform can be computed
exactly:
\begin{eqnarray}
P(Q_\tau) =  {1 \over 4}\exp \left( \frac{\sigma_W ^2}{2}\right) \left[ e^{Q_\tau - \langle Q_\tau \rangle}
{\rm erfc} \left(\frac{Q_\tau-\langle Q_\tau \rangle+\sigma_W^2}{\sqrt{2\sigma_W^2}}\right) +
\nonumber \right.\\ \left.
e^{-(Q_\tau - \langle Q_\tau \rangle)}{\rm erfc} \left(\frac{-Q_\tau+\langle Q_\tau \rangle+\sigma_W^2}
{\sqrt{2\sigma_W^2}}\right)\right] \,,
\label{PDFq}
\end{eqnarray}
where erfc stands for the complementary Erf function.
In Fig.~\ref{fig:sinusFT}c, we have plotted the analytical PDF from eq.~(\ref{PDFq}) together with the experimental ones,
using values of $\sigma_W^2$ and $\langle Q_\tau \rangle$ from the experiment.
The agreement is perfect for all values of $n$ and with no adjustable parameters.
Using eq.~(\ref{PDFq}), we isolate three different regions for $S(q_\tau)$:

(I) if $Q_\tau > \sigma_{W}^2 + |\langle Q_\tau \rangle|= 3 |\langle Q_\tau \rangle| + {\cal O}(1)$, then
$S(q_\tau) = 2+{\cal O}({1 \over \tau})$.
This case corresponds to large fluctuations and the PDF can be pictured as exponential with a non-vanishing mean.

(II) if $Q_\tau < \sigma_{W}^2 - |\langle Q_\tau \rangle| = |\langle Q_\tau \rangle| + {\cal O}(1)$, then
$S(q_\tau)=\Sigma(n) q_\tau + {\cal O}({1 \over \tau})$ with $\Sigma_q(n) = \frac{2 |\langle Q_\tau \rangle|}{\sigma_W^2} = \Sigma_w(n) $. In this case, heat fluctuations are small and behave like work fluctuations. The slope $\Sigma(\tau)$ is the same as the one found in the case of work fluctuations.
The exact correction to the asymptotic value is plotted in Fig.~\ref{fig:sinusFT}e and again
it matches perfectly the experimental behavior.

(III) for $\sigma_{W}^2 - |\langle Q_\tau \rangle|  < Q_\tau < \sigma_{W}^2+ |\langle Q_\tau \rangle |$,
there is an intermediate region connecting cases (I) and (II) by a second order polynomial
$S(q_\tau)=2. - {\Sigma (\tau) \over 4}(q_\tau-(1 + {2 \over \Sigma (\tau)}))^2+{\cal O}({1 \over \tau})$.

Those three regions offer a perfect description of the three domains observed experimentally (Fig.~\ref{fig:sinusFT}d).

Now, we examine the limit of infinite $\tau$ in which SSFT is supposed to hold but which depends
on the variable we use : either the heat $Q_\tau$ or the normalized heat $q_\tau$.
First we discuss $Q_\tau$. The asymptotic behavior of the PDF of $Q_\tau$ (eq.~(\ref{PDFq})) for large $\tau$ is Gaussian with variance $\sigma_W^2$. Thus, the PDF of $Q_\tau$ coincides with the PDF of $W_\tau$ for $\tau$ strictly infinite. As we have already shown, work fluctuations satisfy the conventional SSFT, therefore heat fluctuations also satisfy the conventional SSFT (eq.~(\ref{FT})).
We have found three different regions defined by two characteristic values, but
in the limit of infinite time $\tau$, only region (II) is relevant: region (II) is bounded
from above by $|\langle Q_\tau \rangle| + {\cal O}(1)$ with the average $\langle Q_\tau \rangle$
being linear in $\tau$. We see that all the behavior of the fluctuations of $Q_\tau$
is captured by region (II) where $S(q_\tau)$ is linear for all $Q_\tau$ and SSFT holds.
Second, we turn to the normalized heat $q_\tau$. As the average value of $Q_\tau$ is
linear with $\tau$, rescaling by $\langle Q_\tau \rangle$ is equivalent to a division by $\tau$ ;  the mean of $q_\tau$ is then $1$. This normalization changes into constants the two characteristic
values which delimit the three regions : the boundary between (II) and (III) is now $1+{\cal O}(1/\tau)$ and the boundary between (III) and (I) is $3+{\cal O}(1/\tau)$.
The function $S(q_\tau)$ is not linear for large values of $q_\tau$ but only in region (II) ($q_\tau <1$), for small fluctuations.
So SSFT is satisfied only for small fluctuations but not for all values of $q_\tau$, and instead of a FT, we rather have
a fluctuation relation.
These two descriptions in terms of $Q_\tau$ or $q_\tau$ are in fact a problem of two non-commutative limits.
The first description implies that one takes the limit $\tau$ infinite before taking the limit of large $Q_\tau$.
The second description does the opposite.
However, the probability to have large fluctuations decreases with $\tau$ and experimentally,
for large $\tau$, only the region (II) where SSFT holds ($q_\tau<1$) can be seen.

\begin{figure}
\centerline{\includegraphics[width=0.7\linewidth]{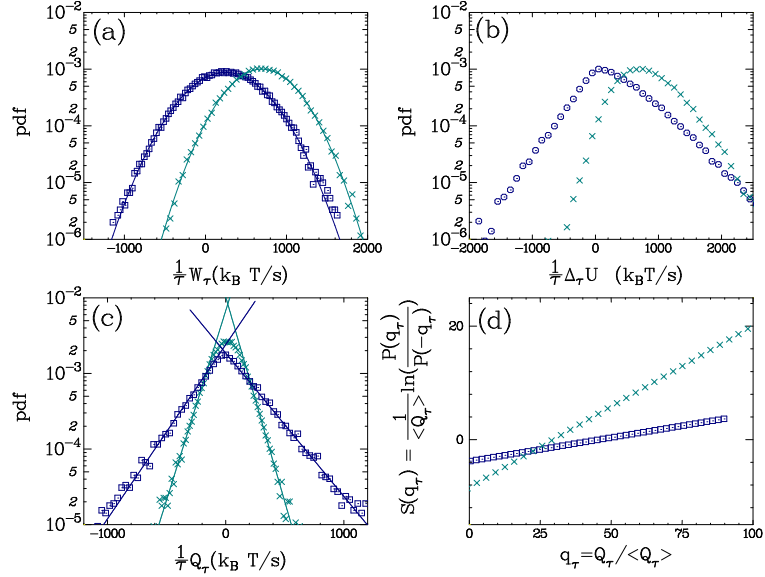}}
\caption{TFT. All plots are given for two values of $\tau/\tau_\alpha$: 0.3 $(\Box)$ and 1.04 $(\times)$.
   a) PDFs of ${1 \over \tau}W_\tau$. Continuous lines are Gaussian fits. b) PDFs of ${1 \over \tau}\Delta U_\tau$. c) PDFs of ${1 \over \tau}Q_\tau$. Continuous lines are exponential fits for large $Q_\tau$. d) Functions $S(q_\tau)$ for large heat fluctuations $q_\tau$ computed using exponential fits.}
\label{fig:TFT}
\end{figure}

Finally, we briefly report results for the Transient Fluctuation Theorem.
For this, we choose a torque $M(t) = M_o t/\tau_r$ linear in time, and
the system is at equilibrium at $t_i=0$ ($M(t_i=0) = 0$ pN.m).
Unlike with the periodic driving, the average of the variation of internal energy
$\langle \Delta U_\tau \rangle$ is not vanishing. The work done by $M(t)$ is used
by the system to increase his internal energy but a small amount of energy is lost
by viscous dissipation and exchange with the thermostat. The PDF of $W_\tau/\tau$,
$\Delta U_\tau/\tau$ and $Q_\tau/\tau$ are plotted in Fig.~\ref{fig:TFT}
for different values of $\tau/\tau_\alpha$. Averages $\langle W_\tau/\tau \rangle$ and
$\langle \Delta U_\tau/\tau \rangle$ are linear in $\tau$.
However, their difference (eq.~(\ref{Qdef})) $\langle Q_\tau/\tau \rangle$
is constant and of the order of a few $k_B T/s$ (Fig.~\ref{fig:TFT}c).
The shapes of the PDF are different from the ones obtained with the periodic forcing.
Work fluctuations have a Gaussian PDF for any values of $\tau$, moreover TFT holds for
$W_\tau$~\cite{Douarche2006}. In contrast $\Delta U_\tau$ have a different probability
distribution. Fig.~\ref{fig:TFT}b shows that the PDF are not symmetric around the mean value.
Extreme events have again an exponential distribution. For exactly the same reasons as in
the case of a periodic forcing, the PDFs of $Q_\tau/\tau$ are not Gaussian and have
exponential tails for extreme fluctuations : $P(Q_\tau/\tau) = A \exp (-\alpha |Q_\tau/\tau|)$. As we can see in fig.~\ref{fig:TFT}c, the PDFs are not symmetric around the mean value, thus there are two pairs ($\alpha$, $A$), one for positive value of large fluctuations ($\alpha_+$, $A_+$) and one for negative ($\alpha_-$, $A_-$). Thus, there is a simple expression of $S(q_\tau)$ for large fluctuations:
\begin{equation}
S(q_\tau) = (\alpha_+ - \alpha_-) q_\tau + {1 \over \langle Q_\tau \rangle} \ln \left( \frac{A_+}{A_-} \right)
\label{TFTlargefluctuation}
\end{equation}
It can be seen experimentally in fig.~\ref{fig:TFT}d. $S(q_\tau)$
is not proportional to  $q_\tau$, therefore TFT is not satisfied
for finite time. However, for large value of $\tau$, the PDF of
$Q_\tau$ becomes symmetric around the mean value and only the Gaussian behavior remains. Thus, TFT appears to be satisfied
experimentally in the limit of infinite $\tau$. This breaks the
expected property of TFT to be valid at any time.

In conclusion we have proposed an SSFT for heat fluctuations of a harmonic
oscillator driven in a stationary out of equilibrium state by a periodic external force.
An exact expression of the PDFs of the heat $Q_\tau$ averaged over a time $\tau$
is given. This PDF is in perfect agreement with experimental data.
For finite times, we have isolated different behaviors: one for small fluctuations
(Gaussian behavior) and the other for extreme fluctuations (exponential behavior).
SSFT holds for infinite time.
We have also studied a TFT for $Q_\tau$ using linear forcing and found that FT is satisfied
only in the limit of large times.

We thank G. Gallavotti for useful discussions. This work has been partially supported by ANR-05-BLAN-0105-01.

\end{document}